\documentclass[%
reprint,
superscriptaddress,
amsmath,amssymb,
aps,
]{revtex4-2}

\usepackage{graphicx}% Include figure files
\usepackage{dcolumn}% Align table columns on decimal point
\usepackage{bm}% bold math
\usepackage{mathrsfs}
\usepackage{float}
\usepackage{physics}
\usepackage{siunitx}
\usepackage[colorlinks,linkcolor=blue,anchorcolor=blue,urlcolor=blue,citecolor=blue]{hyperref}
\begin{document}

\preprint{APS/123-QED}

\title{Thouless pumping and generation of squeezed Fock-state superpositions in a Fock-state lattice}% Force line breaks with \\

\author{Ruo Kun Cai}
\thanks{These authors contributed equally to this work.}
\affiliation{College of Science, National University of Defense Technology, Changsha, Hunan 410073, China\\}

\author{Ling Lin}
\thanks{These authors contributed equally to this work.}
\affiliation{Institute of Quantum Precision Measurement, State Key Laboratory of Radio Frequency Heterogeneous Integration, College of Physics and Optoelectronic Engineering, Shenzhen University, Shenzhen 518060, China}

\author{Chun Wang Wu}
\affiliation{College of Science, National University of Defense Technology, Changsha, Hunan 410073, China\\}
\affiliation{Hunan Key Laboratory of Mechanism and Technology of Quantum Information, Changsha, Hunan 410073, China\\}

\author{Zhi Jiao Deng}
\email{dengzhijiao926@hotmail.com}
\affiliation{College of Science, National University of Defense Technology, Changsha, Hunan 410073, China\\}
\affiliation{Hunan Key Laboratory of Mechanism and Technology of Quantum Information, Changsha, Hunan 410073, China\\} 

\author{Ping Xing Chen}
\affiliation{College of Science, National University of Defense Technology, Changsha, Hunan 410073, China\\}
\affiliation{Hunan Key Laboratory of Mechanism and Technology of Quantum Information, Changsha, Hunan 410073, China\\}

\date{\today}% It is always \today, today,

\begin{abstract}
In this paper, Thouless pumping in a one-dimensional semi-infinite Fock-state lattice is investigated. A distinctive feature of such lattices is the intrinsic $\sqrt{n}$-dependent coupling arising from the bosonic mode, which leads to spatially nonuniform hopping amplitudes. In the dimer limit, the topological invariants and the quantized transport dynamics in the Fock-state basis are numerically evaluated and analyzed. By introducing an additional inter-cell coupling and applying a squeezing transformation, the framework is then extended to Thouless pumping in the squeezed Fock-state basis, where a topologically protected scheme for preparing superpositions of squeezed Fock states is proposed. This study establishes Thouless pumping in Fock-state lattices as a useful tool for quantum state engineering, shifting the focus from observing topological transport to harnessing it for the preparation of non-classical states of the bosonic mode.
\end{abstract}

\maketitle

%\tableofcontents

\section{\label{sec:level1}Introduction}
Since Thouless’s seminal proposal in 1983 \cite{Thouless1983}, topological pumping has emerged as one of the most paradigmatic manifestations of topology in quantum systems. A Thouless pump generates a quantized particle current through slow and periodic modulation of system parameters in the absence of any external bias \cite{CitroAidelsburger2023}. By interpreting time as an extra dimension, a spatially one-dimensional Thouless pump is equivalent to a two-dimensional quantum Hall system. Consequently, the amount of particle transported per cycle is determined by a topological invariant—the Chern number—and remains quantized against weak disorder and many-body interactions \cite{NiuThouless1984}. This topological robustness, rooted in the geometric Berry phase, has made Thouless pumping a fertile ground for exploring quantum sensing \cite{Sensing}, quantum computation \cite{Splitting} and on-chip photonic devices \cite{Onchip}. Over the past decades, Thouless pumping has been successfully realized in a variety of experimental platforms, including ultracold atoms in optical superlattices \cite{Coldatom1, Coldatom2}, photonic waveguide arrays \cite{Optics}, and mechanical systems \cite{Mechanic}. Meanwhile, current research has been pushing the framework in several directions, including the roles of interactions \cite{Interaction1, Interaction2, Interaction3, Interaction4}, nonlinearity \cite{Nonlinear1, Nonlinear2, Nonlinear3}, and nonadiabatic effects \cite{Nonadiabatic}, as well as topological pumping in higher dimensions \cite{Highdimension1, Highdimension2}. 

On the other hand, synthetic dimensions \cite{Synthetic1, Synthetic2}, using internal states or angular momentum as artificial dimensions, enable the simulation of topological models with enhanced dimension or large lattice number. Among various implementations of synthetic dimensions, the Fock-state lattice (FSL), where Fock states of a bosonic mode serve as lattice sites, provides a particularly natural platform \cite{fock1, fock2, fock3}. FSL has been used to simulate canonical topological models such as the Su-Schrieffer-Heeger (SSH) model \cite{SSH1, SSH2, SSH3}, the Haldane model \cite{SSH1}, synthetic gauge fields \cite{Flux1, Flux2}, and non-Hermitian skin effects \cite{Skin1, Skin2}. The intrinsic $\sqrt{n}$ dependence of the hopping amplitude, originating from the bosonic nature of the Fock states with $n$ labeling the excitation number of the mode, breaks translational symmetry and produces new effects, including bound states \cite{SSH2,SSH3} and spatially confined skin effect \cite{Skin2}. Yet the question of Thouless pumping in this setting remains entirely open. Specifically, the intrinsic $\sqrt{n}$ dependence, which fundamentally distinguishes the FSL from uniform lattices, raises three interrelated issues: whether quantized pumping survives this inhomogeneous coupling, how to characterize its topological invariant in such a nonuniform system, and what new physical phenomena might emerge from the interplay between the nonuniform hopping and the adiabatic cyclic modulation. 

In this work, we construct a nonuniformly coupled Rice–Mele model on a one-dimensional semi-infinite Fock-state lattice and study its topological pumping. 
This model describes a single two-level system interacting with a bosonic mode. Owing to the intrinsic $\sqrt{n}$-dependent hopping that breaks translational symmetry, well-defined Wannier wave packets suitable for quantized transport cannot be constructed under generic modulations.
However, in the dimer limit, the nonuniform coupling does not hinder pumping, and the process becomes equivalent to the uniform case, as verified by the real-space calculation of the Chern number \cite{Thouless1983, Xoperator}. By introducing an extra intercell coupling, we obtain an anisotropic Rabi model \cite{ARabi} that enables topological pumping in the basis of squeezed Fock states. This effect originates solely from the $\sqrt{n}$ dependence of the hopping amplitudes inherent to bosonic modes and is absent in uniform coupling models. The pumping provides a robust, topologically protected method for preparing superpositions of squeezed Fock states. Such superposition states can be utilized in quantum error correction codes \cite{Errorcorrection}, making our scheme a promising tool for bosonic quantum information processing.

The rest of this paper is structured as follows. In Sec. II, we present the Rice–Mele model on the Fock-state lattice, compute the Chern number in the dimer limit, and demonstrate the quantized transport of Fock states via numerical simulations. In Sec. III, we introduce an extra intercell coupling to examine quantized pumping in the squeezed-Fock basis and propose a topologically robust scheme for preparing superpositions of squeezed Fock states based on this mechanism. A brief conclusion is given in Sec. IV.

\section{Model and topological pumping characterization} 
\begin{figure}[t]
	\includegraphics[width=0.48\textwidth]{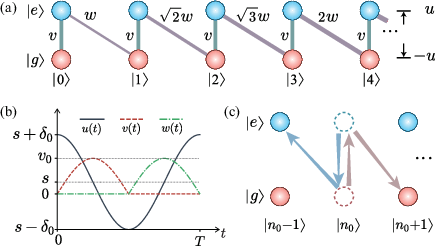}% Here is how to import EPS art
	\centering
	\caption{\label{figure1}Schematic of the Rice–Mele model realized on the Fock-state lattice. (a) In each unit cell labeled by $n$, the two sublattice sites are represented by $|g,n\rangle$ and $|e,n\rangle$. The nearest-neighbor hoppings are $v$ within each unit cell and $\sqrt{n}w$ between adjacent cells, and the sublattice sites are detuned by the staggered potential $\pm u$. (b) Periodic driving protocol of the three parameters $u(t)$, $v(t)$, and $w(t)$ over one pump cycle. (c) Adiabatic transport in the dimer limit. Starting from either internal state, the particle is transported by one unit cell per cycle, with the direction determined by the initial internal state.}
\end{figure}

\begin{figure}[t]
	\includegraphics[width=0.48\textwidth]{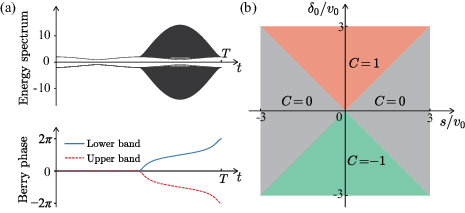}% Here is how to import EPS art
	\centering
	\caption{\label{figure2}Topological characterization of the Fock-state lattice pump. (a) Instantaneous energy spectrum (upper panel) and Berry phase (lower panel) over one pumping cycle. The numerical calculation is performed with a cutoff at $n=201$ under periodic boundary conditions. Parameters: $\delta_0=2$, $s=0$, $v_0=1$. (b) Phase diagram of the lower-band Chern number in the $(s/v_0,\delta_0/v_0)$ parameter space. The phase boundary $|s|=|\delta_0|$ separates the topological regime with $C=\pm 1$ from the trivial regime with $C=0$.}
\end{figure}

\begin{figure}[t]
	\includegraphics[width=0.48\textwidth]{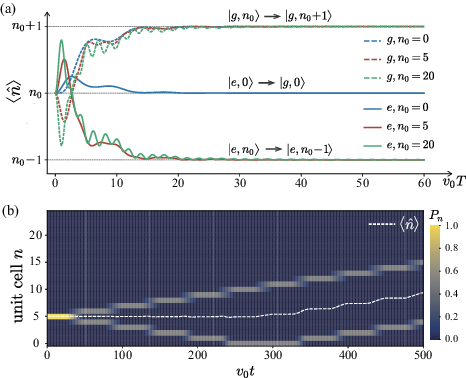}% Here is how to import EPS art
	\centering
	\caption{\label{figure3}Topological pumping dynamics in the Fock-state lattice. (a) Mean position $\langle \hat{n} \rangle = \langle \hat{a}^\dagger \hat{a} \rangle$ as a function of the pumping period $T$ after a single pumping cycle, for different initial states $\ket{g,n_0}$ or $\ket{e,n_0}$ with $n_0 = 0, 5, 20$. (b) Probability distribution $P_{n}$ across each unit cell as a function of the evolution time, starting from the initial superposition $\frac{1}{\sqrt{2}}(\ket{g,5}+\ket{e,5})$, with $v_0 T = 50$; the white dashed line denotes the time evolution of the mean position. Parameters: $\delta_0 = 2$, $s = 0$, $v_0 = 1$.}
\end{figure}

A minimal model for Thouless pumping is the Rice–Mele model, which extends the SSH model by introducing a staggered on-site potential and applying time-periodic modulations to all parameters \cite{RiceMele}. The SSH model has been implemented on Fock-state lattices in several recent works \cite{SSH3, Skin2}. Consider a spin-boson coupled system consisting of a two-level atom and a bosonic mode, the central idea is to identify all states $|g,n\rangle$ and $|e,n\rangle$ of this system as lattice sites, with the spin index labeling the sublattice and the bosonic excitation number $n$ labeling the unit cell along a semi-infinite synthetic chain. With the coupling configuration and on-site potentials depicted in  Fig.~\ref{figure1}(a), the Hamiltonian of the system reads
\begin{equation}
	\label{ham}
	\hat{H}_{\rm JC}(t) = u(t)\,\hat{\sigma}_z + v(t)\,\hat{\sigma}_x + w(t)\left(\hat{\sigma}_- \hat{a}^\dagger + \hat{\sigma}_+ \hat{a}\right),
\end{equation}
here $\hat{\sigma}_z = |e\rangle\langle e| - |g\rangle\langle g|$ and $\hat{\sigma}_x = |e\rangle\langle g| + |g\rangle\langle e|$ are the Pauli operators on the two internal states $|g\rangle$ and $|e\rangle$, with $\hat{\sigma}_- = |g\rangle\langle e|$ and $\hat{\sigma}_+ = |e\rangle\langle g|$ the corresponding lowering and raising operators, and $\hat{a}$, $\hat{a}^\dagger$ the annihilation and creation operators for the bosonic mode. The coefficients $u(t)$, $v(t)$, and $w(t)$ denote the time-periodic staggered on-site potential, the intracell hopping, and the intercell hopping, respectively. The last term in Eq.~(\ref{ham}) describes the Jaynes--Cummings (JC) coupling between the two-level system and the bosonic mode \cite{JCmodel}.

As the intercell hopping carries the intrinsic $\sqrt{n}$ factor that breaks translational symmetry  in the Fock state lattice, Wannier states are not well defined here. As a result, under generic modulations we find no suitable initial wave packet that exhibits quantized transport. Therefore, the modulations are restricted to the dimer limit, in which $v(t)$ and $w(t)$ are nonzero in separate halves of the pumping cycle, as illustrated in Fig.~\ref{figure1}(b)
\begin{equation}
	\label{modulation}
	\begin{split}
		u(t) &= \delta_0 \cos\left(\frac{2\pi t}{T}\right) + s,\\
		v(t) &=
		\begin{cases}
			v_0 \sin\left(\frac{2\pi t}{T}\right), & 0 \le t < \frac{T}{2},\\
			0, & \frac{T}{2} \le t < T,
		\end{cases}\\
		w(t) &=
		\begin{cases}
			0, & 0 \le t < \frac{T}{2},\\
			- v_0 \sin\left(\frac{2\pi t}{T}\right), & \frac{T}{2} \le t < T,
		\end{cases}
	\end{split}
\end{equation}
where $\delta_0$ and $v_0$ are the modulation amplitudes for the on-site and hopping terms, respectively, $s$ is a constant offset of the on-site potential, and $T$ is the time period. During the first half of the cycle, the dynamics transfer the population within the same unit cell, while in the second half, the population is transferred between neighboring unit cells, as illustrated in Fig.~\ref{figure1}(c). Starting from the site $|g,n_0\rangle$, the population sequentially transfers to $|e,n_0\rangle$ and then to $|g,n_0+1\rangle$. Thus, for a particle initially in the state $\ket{g,n_0}$, it is transported to the right by exactly one unit cell over a full cycle. Conversely, if the initial state is $\ket{e,n_0}$, it is transported to the left by one unit cell. These features are consistent with the standard Rice--Mele pump and demonstrate a quantized, state-dependent directional transport. 

The robustness of topological transport is guaranteed by the quantized Chern number, which is given by the net change of polarization over one pumping cycle: $C = \Delta P = P(T) - P(0)$ \cite{Thouless1983}. The polarization is the Berry phase divided by $2\pi$ \cite{RevModPhys.66.899}. In translation-invariant systems, this phase is obtained from the integral of the Berry connection over the Brillouin zone \cite{Zak1989}. In the present nonuniform Fock-state lattice, however, the polarization can be evaluated through the Resta position operator \cite{Xoperator} $\hat{X}_{\mathrm{op}} = \sum_{n} e^{i 2\pi n/N} \left( |g,n\rangle\langle g,n| + |e,n\rangle\langle e,n| \right)$, where $N$ is the number of unit cells. 
Projecting this operator onto the subspace spanned by a chosen energy band yields the overlap matrix $ {{\cal U}(t)}$, where ${\left[ {{\cal U}(t)} \right]_{n,m}} = \langle {\psi _n}|{{\hat X}_{{\rm{op}}}}|{\psi _m}\rangle $ with eigenstates $|\psi_{n}\rangle$ and $|\psi_{m}\rangle$ belonging to the chosen energy band, and the polarization is then obtained from the argument of its determinant as $P(t) = \frac{1}{2\pi}\mathrm{Arg}\left(\det \mathcal{U}(t)\right)$ \cite{BiancoResta2011}. 
This real-space approach has been used to compute polarization and topological invariants in disordered and other nonuniform systems \cite{Disorder, Quasi}, where the momentum-space formulation is no longer applicable.

Fig. \ref{figure2}(a) displays the instantaneous energy spectrum and the time evolution of the Berry phase over one pumping cycle. 
Given that the system contains infinitely many lattice sites, we truncate the system to a sufficiently large number of sites and impose periodic boundary conditions in the numerical calculation, which reduces the effect of boundary states and enhances the stability of numerical results.
 In the first half of the cycle, the spectrum is highly degenerate; in the second half, this degeneracy is lifted by the $\sqrt{n}$ factor, while the upper and lower bands remain symmetric about $E = 0$. At the initial time, with $u > 0$, the instantaneous eigenstates of the lower and upper bands are $|g,n\rangle$ and $|e,n\rangle$, respectively. Over one complete pumping cycle, the Berry phase winds by $+2\pi$ for the lower band and $-2\pi$ for the upper band, corresponding to Chern numbers $C = +1$ and $C = -1$, respectively. These results agree with the analysis above: a particle initially prepared in $|g,n\rangle$ is transported to the right by one unit cell, while one initially in $|e,n\rangle$ is transported to the left by the same amount. 

To determine the parameter regimes in which this quantized pump persists, the Chern number is evaluated across the $(s/v_0,\delta_0/v_0)$ parameter space. The resulting phase diagram for the lower band is shown in Fig. \ref{figure2}(b). The topological phase transition occurs when the pump path passes through the gap-closing point at $(u, v, w) = (0, 0, 0)$, where the instantaneous energy gap vanishes. This condition determines the phase boundary $|s| = |\delta_0|$, across which the Chern number changes from $C = \pm 1$ to $C = 0$. The Chern numbers of the upper and lower bands sum to zero, as expected for a two-band model. Moreover, reversing the sign of $\delta_0$ effectively exchanges the roles of the upper and lower bands, which flips the sign of the Chern number for each band. 

Topological pumping relies on adiabatic parameter tuning, which requires that the time scale of the variation be much larger than the inverse gap ($\hbar/\Delta E$) to suppress non-adiabatic transitions. Figure \ref{figure3}(a) plots, as a function of the cycle period $T$, the mean position $\langle \hat{n} \rangle = \langle \hat{a}^\dagger \hat{a} \rangle$ after one complete pumping cycle for different initial states prepared in the lower band $\ket{g,n}$ or the upper band $\ket{e,n}$. As $T$ increases, the mean positions converge to stable values: sites in the lower and upper bands shift by one unit cell in the directions of increasing and decreasing $n$, respectively. The only exception is the boundary site $\ket{e,0}$, whose mean position remains unchanged, although it undergoes an interband transfer from the upper band $\ket{e,0}$ to the lower band $\ket{g,0}$. Overall, longer periods yield better adiabatic transport, whereas oscillations appear for intermediate $T$ values, arising from interference effects due to residual non-adiabatic transitions. Furthermore, exploiting this interband transfer at the boundary site, we can engineer superposition states of Fock states. As shown in Fig.~\ref{figure3}(b), the initial superposition $\frac{1}{\sqrt{2}}(\ket{g,5}+\ket{e,5})$ evolves, after 10 pumping cycles, into $\frac{1}{\sqrt{2}}(\ket{g,15}+\ket{g,4})$.

\section{Thouless pumping in the squeezed Fock-state basis for state generation}
\begin{figure}[t]
	\includegraphics[width=0.48\textwidth]{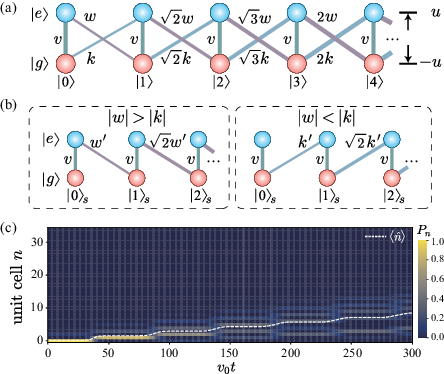}% Here is how to import EPS art
	\centering
	\caption{\label{figure4}Thouless pumping in the squeezed Fock-state basis. (a) Lattice structure in the original Fock-state basis, where both JC and anti-JC couplings coexist. (b) Lattice model in the squeezed Fock-state basis $\ket{n}_s$ after the squeezing transformation, with opposite topological Chern numbers for the two parameter regimes $\abs{w}>\abs{k}$ and $\abs{w}<\abs{k}$. (c) Probability distribution $P_{n}$ across each unit cell in the $\hat{a}$-basis over the evolution time, with the initial state $\ket{g}\otimes\ket{0}_s$, where the bosonic mode is prepared in a squeezed vacuum state; the white dashed line marks the trajectory of the mean position $\langle \hat{a}^\dagger \hat{a} \rangle$. Parameters: $v_0 T = 50$, $k/w = 0.4$, $\delta_0 = 2$, $s = 0$, $v_0 = 1$.}
\end{figure}

\begin{figure}[t]
	\includegraphics[width=0.48\textwidth]{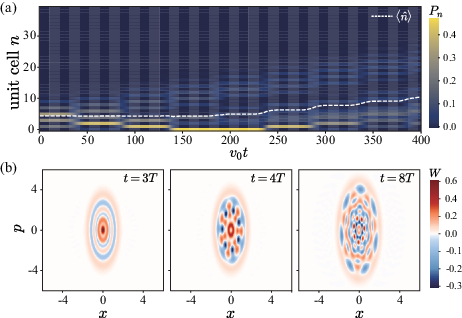}% Here is how to import EPS art
	\centering
	\caption{\label{figure5}Generation of squeezed Fock state superpositions via Thouless pumping in the squeezed Fock-state basis. (a) Probability distribution across each unit cell in the $\hat{a}$-basis over the evolution time, starting from the initial state $\frac{1}{\sqrt{2}}(\ket{e}+\ket{g})\otimes\ket{3}_s$; the white dashed line denotes the time evolution of the mean position $\langle \hat{a}^\dagger \hat{a} \rangle$. In each pumping cycle, the squeezed Fock state $\ket{n}_s$ undergoes $n \to n+1$ and $n \to n-1$ for the components associated with $\ket{g}$ and $\ket{e}$, respectively; after the interband transfer at the boundary, both components follow $n \to n+1$ per cycle. (b) The corresponding Wigner functions $W(x,p)$ of the bosonic mode at $t = 3T$, $4T$, and $8T$, which evolve from a squeezed concentric ring to interference fringes, indicating the transition of the bosonic mode from a mixed state to a superposition of two squeezed Fock states. Other parameters are the same as in Fig.~\ref{figure4}(c).}
\end{figure}

In the preceding section, the topological transport within a lattice whose sites are spanned by Fock states has been discussed. A key feature is that particles on the two sublattices, $\ket{g,n}$ and $\ket{e,n}$, exhibit opposite propagation directions. 
 We now extend the Hamiltonian in Eq.~(\ref{ham}) by introducing an anti-Jaynes--Cummings (anti-JC) coupling term, such that the combined JC and anti-JC couplings induce a squeezing effect in the system \cite{antiJC1, antiJC2, antiJC3}. Specifically, as illustrated in Fig.~\ref{figure4}(a), a coupling between $\ket{g,n}$ and $\ket{e,n+1}$ in neighboring unit cells is added. The modified system Hamiltonian is then given by
\begin{equation}
	\label{ham_rabi}
	\begin{split}
		\hat{H}_{\rm Rabi}(t) ={}& u(t)\,\hat{\sigma}_z + v(t)\,\hat{\sigma}_x + w(t)\left(\hat{\sigma}_- \hat{a}^\dagger + \hat{\sigma}_+ \hat{a}\right) \\
		&+ k(t)\left(\hat{\sigma}_- \hat{a} + \hat{\sigma}_+ \hat{a}^\dagger\right),
	\end{split}
\end{equation}
where the last term represents the anti-JC coupling, with $k(t)$ denoting its strength. For $w(t) \neq k(t)$, the Hamiltonian $\hat{H}_{\rm Rabi}(t)$ is known as the anisotropic Rabi model \cite{ARabi}, which has been implemented in various platforms including trapped ions and superconducting circuits \cite{antiJC2, Song2026}.

To exploit the squeezing effect, we require a ratio $k(t)/w(t) = \text{const.}$ and introduce a new bosonic operator $\hat{b}$ via the Bogoliubov transformation $\hat{a} = \hat{b}\cosh r + \hat{b}^\dagger\sinh r$. By choosing the squeezing parameter $r$ appropriately, one of the two coupling channels can be eliminated. For $\abs{w} > \abs{k}$, choosing $\tanh r = -k/w$ eliminates the anti-JC term, yielding a pure JC model:
\begin{equation}
	\label{jc_result}
	\hat{H}_{\rm Rabi}(t) = u(t)\hat{\sigma}_z + v(t)\hat{\sigma}_x + w'(t)\left(\hat{\sigma}_- \hat{b}^\dagger + \hat{\sigma}_+ \hat{b}\right),
\end{equation}
with $w'(t) = \sqrt{w^2(t) - k^2(t)}$. Conversely, for $\abs{w} < \abs{k}$, choosing $\tanh r = -w/k$ eliminates the JC term, yielding a pure anti-JC model:
\begin{equation}
	\label{anti_jc_result}
	\hat{H}_{\rm Rabi}(t) = u(t)\hat{\sigma}_z + v(t)\hat{\sigma}_x + k'(t)\left(\hat{\sigma}_- \hat{b} + \hat{\sigma}_+ \hat{b}^\dagger\right),
\end{equation}
with $k'(t) = \sqrt{k^2(t) - w^2(t)}$. In either case, the original Hamiltonian is recast into a simpler form containing only a single coupling channel.

The $\hat{b}$-basis introduced above is in fact a squeezed Fock-state basis in the original $\hat{a}$-representation. To see this, we invert the Bogoliubov transformation:
\begin{equation}
	\hat{b} = \hat{a}\cosh r - \hat{a}^\dagger\sinh r,
\end{equation}
which is equivalently expressed as $\hat{b} = \hat{S}^\dagger(r)\hat{a}\hat{S}(r)$ with the squeezing operator $\hat{S}(r) = \exp[\frac{r}{2}(\hat{a}^2 - \hat{a}^{\dagger 2})]$. Consequently, the creation operators transform as $\hat{b}^\dagger = \hat{S}^\dagger(r)\hat{a}^\dagger\hat{S}(r)$, and the vacuum states satisfy $|0\rangle_b = \hat{S}^\dagger(r)|0\rangle_a$. Substituting these into the definition $|n\rangle_b = (\hat{b}^\dagger)^n/\sqrt{n!}\,|0\rangle_b$ yields the Fock-state relation
\begin{equation}
	\ket{n}_b = \hat{S}^\dagger(r)\ket{n}_a.
\end{equation}
This confirms the identification $\ket{n}_a \equiv \ket{n}$ and $\ket{n}_b \equiv \ket{n}_s$, indicating the squeezed Fock state. The simplified Hamiltonians in Eqs.~(\ref{jc_result}) and (\ref{anti_jc_result}) therefore describe lattice models in this squeezed basis, with opposite topological Chern numbers, as shown in Fig.~\ref{figure4}(b).
Consequently, the topological analysis presented in the previous section applies directly.

For a squeezed Fock state $\ket{n}_s$, over one adiabatic modulation cycle, the quantum number $n$ changes by $\pm 1$, i.e., $\ket{n}_s \to \ket{n \pm 1}_s$. To relate this dynamics to the original Fock basis, we examine the change in the average position between adjacent squeezed Fock states. Using the relation between $\hat{a}$ and $\hat{b}$, this yields
\begin{equation}
	{}_s\langle n \pm 1| \, \hat{a}^\dagger \hat{a} \, |n \pm 1\rangle_s - {}_s\langle n| \, \hat{a}^\dagger \hat{a} \, |n\rangle_s = \pm \cosh 2r.
\end{equation}
Thus, in the original Fock basis, the steps in the average position remain equally spaced, with the step size multiplied by a factor of $\cosh 2r > 1$ compared to the original Fock basis. Figure~\ref{figure4}(c) demonstrates the quantized transport process starting from the initial state $\ket{g} \otimes \ket{0}_s$, with the bosonic mode prepared in a squeezed vacuum state. A clear signature of the squeezed Fock-state nature is observed: for even $n$, the state $\ket{n}_s$ exhibits support only on even Fock states of the original $\hat{a}$-mode; similarly, for odd $n$, only odd Fock states are populated. The average position nevertheless maintains quantized plateaus with steps of equal spacing.

Similarly, the opposite transport directions of different bands can be exploited to prepare superpositions of squeezed Fock states. As illustrated in Fig.~\ref{figure5}(a), the system is initially prepared in the state $\frac{1}{\sqrt{2}}(\ket{e}+\ket{g})\otimes\ket{3}_s$ and then undergoes eight pumping cycles. The quantum numbers $n$ of the squeezed Fock states linked to $\ket{g}$ and $\ket{e}$ increase and decrease by $1$ per cycle, respectively. Notably, the boundary state $\ket{e}\otimes\ket{0}_s$ behaves differently: it undergoes an interband transfer from the upper to the lower band $\ket{g}\otimes\ket{0}_s$ after one complete cycle, and then reverses direction toward increasing $n$ in subsequent cycles. Before this interband transfer, the system resides in the entangled state $\frac{1}{\sqrt{2}}(\ket{e}\otimes\ket{3-m}_s + \ket{g}\otimes\ket{3+m}_s)$ ($m \le 3$), where $m$ is the number of pumping cycles. In this entangled regime, the bosonic mode is in a mixed state of two squeezed Fock states, and the corresponding Wigner function shows a squeezed concentric ring. Once this interband transfer is done, the state becomes the product state $\frac{1}{\sqrt{2}}\ket{g}\otimes(\ket{m-4}_s + \ket{3+m}_s)$ ($m > 3$), where the bosonic mode is instead in a superposition of two squeezed Fock states, yielding interference fringes in the Wigner function, as shown in Fig.~\ref{figure5}(b). Actually, superpositions of squeezed Fock states have been experimentally realized in trapped-ion systems \cite{antiJC2}, yet via pulsed schemes that require precise control of pulse durations. In contrast, our approach does not rely on such precise timing, and the quantized transport is topologically robust against fluctuations in the adiabatically modulated parameters. Crucially, this scheme relies on the $\sqrt{n}$-dependent coupling structure of the Fock-state lattice; without it, the squeezing transformation would no longer be valid and the method would fail.

\section{Conclusion}
This work presents a study of Thouless pumping in a one-dimensional semi-infinite Fock-state lattice. Under generic parameter modulations, due to the intrinsic $\sqrt{n}$-dependent coupling and the single-particle nature of the model, initial wave-packet states that support quantized transport have not been identified. We therefore restrict our discussion to the dimer limit, where the phase diagram of topological invariants and the quantized transport dynamics are analyzed. The results are then extended to topological transport in the squeezed Fock-state basis, based on which a robust scheme for preparing superpositions of squeezed Fock states is further proposed. These studies demonstrate Thouless pumping in Fock-state lattices as a reliable tool for quantum state manipulation on quantum optical platforms, providing an alternative route and new insights for quantum information processing.

\begin{acknowledgments}
Z. J. Deng is grateful to useful discussion with Jie Zhang. This work was supported by the National Natural Science Foundation of China under Grants No. 12174448, No. 12505025, No. 11574398, and No. 12574403, and by the Innovation Research Foundation of NUDT.
\end{acknowledgments}
    
% whether or not they are actually referenced in the text. This is ap

\bibliography{thouless}% Produces the bibliography via BibTeX.

\end{document}